\documentclass{aa}  

\usepackage{graphicx}
\usepackage[varg]{txfonts}
\usepackage{array}

\usepackage{natbib}

\newcommand{\rtwo}{{r}_{200}}

\newcommand{\Msun}{\rm{M}_\odot}

\newcommand{\Mpc}{\rm{Mpc}}
\newcommand{\scd}{\rm{s}}
\newcommand{\km}{\rm{km}}

\usepackage{color}
\definecolor{green}{rgb}{0,0.5,0}
\definecolor{grey}{rgb}{0.4,0.5,0.7}
\newcommand{\concordance}{combined }

\newcommand{\erratum}[1]{\textcolor{red}{#1}}

\begin{document}
   \title{Mass, velocity anisotropy, and pseudo phase--space density profiles of Abell 2142}

   \subtitle{}

   \author{Munari, E.
          \inst{1}
          \and
          Biviano, A.\inst{2,3}
          \and
          Mamon, G. A.\inst{3}}

   \institute{
     Astronomy Unit, Department of Physics, University of Trieste, via
     Tiepolo 11, I-34131 Trieste, Italy \\ \email{munari@oats.inaf.it}
     \and 
     INAF/Osservatorio Astronomico di Trieste, via Tiepolo 11, I-34131
     Trieste, Italy\\ \email{biviano@oats.inaf.it}
     \and
     Institut d'Astrophysique de Paris (UMR 7095: CNRS \& UPMC), 98 bis
     Bd Arago, F-75014 Paris, France\\ \email{gam@iap.fr}
   }


 
  \abstract
  {}
  {We aim to compute the mass and velocity anisotropy profiles of
    Abell 2142 and, from there, the pseudo phase--space density profile
    $Q(r)$ and the density slope - velocity anisotropy $\beta -
    \gamma$ relation, and then to compare them with theoretical
    expectations. }
   {The mass profiles were obtained by using three techniques
     based on member galaxy kinematics, namely the caustic method, the
     method of Dispersion - Kurtosis, and MAMPOSSt. Through the
     inversion of the Jeans equation, it was possible to compute
     the velocity anisotropy profiles.}
   {The mass profiles, as well as the virial values of mass and
     radius, computed with the different techniques agree
     with one another and with the estimates coming from X-ray and
     weak lensing studies. A \concordance mass profile is obtained by
     averaging the lensing, X-ray, and kinematics determinations. The
     cluster mass profile is well fitted by an NFW profile with $c=4.0
     \pm 0.5$. The population of red and blue galaxies appear to have
     a different velocity anisotropy configuration, since red galaxies are
     almost isotropic, while blue galaxies are radially anisotropic,
     with a weak dependence on radius. The $Q(r)$ profile for the red
     galaxy population agrees with the theoretical results found in
     cosmological simulations, suggesting that any bias, relative to
     the dark matter particles, in velocity dispersion of the red
     component is independent of radius. The $\beta - \gamma$ relation
     for red galaxies matches the theoretical relation only in the
     inner region. The deviations might be due to the use of galaxies
     as tracers of the gravitational potential, unlike the non--collisional
     tracer used in the theoretical relation.}
   {}

   \keywords{
               }

   \maketitle
%

\section{Introduction}
\label{sect: intro}
The measure of the mass of cosmological objects, such as clusters of
galaxies, has proven to be an important tool for cosmological
applications. The mass is not a direct observable, and many techniques
have been developed to infer it by measuring observable
quantities. Two methods that are widely used to infer the mass profile
of galaxy clusters are the X-ray and the lensing techniques. The
former makes use of the observations of the X-ray emission of the hot
intracluster plasma (ICM hereafter). The lensing technique makes use
of the relativistic effect of distortion of the trajectories of light
emitted by distant background galaxies caused by the mass of the
observed cluster. These two methods have some limitations either
way. In the case of X-ray technique, the limitation comes from the
usual assumption that the plasma of the cluster is in hydrostatic
equilibrium, and the cluster approximately spherically symmetric
\citep{ettori2002} with no important recent merger activity
\citep{bohringer2010}. As for the lensing technique, its limitation is
that it only allows computing the projected mass, and this includes
all the line-of-sight (\emph{los}) mass contributions. The
complementarity of the different techniques is a strong advantage for
reliably constraining the mass of a cluster.

In this article, we use another kind of information that comes from
the kinematics of the galaxies belonging to the observed cluster. In
fact, the potential well of the cluster, due to the mass, is the main
driver of the orbital motion of the galaxies, which in the absence of
mutual interactions, can be treated as test particles in the
gravitational potential of the cluster. The kinematics of galaxies
therefore carries the information about the mass content of the
cluster. The motion takes place in a six-dimensional phase space, but
the observations are able to capture only three of these dimensions,
namely two for the position and one for the \emph{los} velocity. This
is one of the most important limitations of a mass estimate via
observation of the kinematics of galaxies. To overcome this problem,
most methods assume spherical symmetry.

A spherically symmetric density profile following the universal
relation provided by \cite*{navarro1996,navarro1997} (NFW hereafter)
has often been adopted in these analyses. Such a profile is
characterized by its ``scale radius'' parameter, which is the radius
where the logarithmic slope of the density profile is equal to
$-2$. With the advent of simulations with increasingly higher
resolution, the universality of the NFW density profile has been
questioned (see e.g. \citealp{navarro2004};
\citealp{vogelsberger2011}; \citealp{ludlow2013}). While the
self-similarity of the density profiles of DM-only haloes may not hold
as well as initially thought, another physical parameter appears to
have a quasi-universal radial profile, the pseudo phase--space density
(PPSD hereafter) $Q(r) = \rho / \sigma^3$, where $\rho$ is the total
matter density profile and $\sigma$ the 3D velocity dispersion of the
tracers of the gravitational potential (\citealp{taylor2001};
\citealp{ludlow2010}). Still, some doubts have been raised about its
universality \citep{ludlow2011}. The use of the radial velocity
dispersion instead of the total one has proven to be a valid and
robust alternative for computing the PPSD, in this case called
$Q_r(r)$. The link between these two formulations of the PPSD is
constrained by the velocity anisotropy (hereafter, anisotropy) of the
system, which plays a non trivial role in shaping the structure of a
system. The density profile and the anisotropy profile are in fact
found to correlate. An empirical relation is provided by
\cite{hansen2006} and \cite{ludlow2011}, linking the logarithmic slope
of the density profile $\gamma = d \ln \rho / d \ln r$ and the
anisotropy $\beta(r) = 1-(\sigma_t/\sigma_r)^2$, where $\sigma_r$ and
$\sigma_t$ are the velocity dispersions of the radial component and of
one of the two tangential components, respectively. Hereafter we refer
to anisotropy as $\beta$ or the equivalent $\sigma_r/\sigma_t = 1 /
\sqrt{1-\beta^2}$. We also denote the relation between anisotropy and
logarithmic slope of the density profile as the $\beta-\gamma$
relation.

In this article, we study \object{Abell 2142} (A2142 hereafter), a
rich galaxy cluster at $z \sim 0.09$. The large number of galaxy
members allows us to derive the total mass profile, to test different
models, as well as to perform dynamical analyses in order to derive
the anisotropy of the orbits of galaxies that allows to compute the
pseudo phase--space density profile and the $\beta-\gamma$
relation. This cluster shows evidence of some recent mergers. In fact,
the X-ray emission appears to have an elliptical morphology elongated
in the north--west south--east direction (\citealp{markevitch2000};
\citealp{akamatsu2011}). The merging scenario is also supported by the
presence of substructures of galaxies lying along the direction of the
cluster elongation, as found in the SZ maps by \cite{umetsu2009},
lensing analysis by \cite{okabe2008}, and analysis of the distribution
of los velocities of \cite{owers2011}. However, after analysing
\emph{XMM-Newton} images to investigate the cold fronts of A2142,
\cite{rossetti2013} argue that the mergers have intermediate mass
ratios rather than major ones.

Throughout this paper, we adopt a $\Lambda CDM$ cosmology with $H_0 =
70\, \km \; \scd^{-1} \; \Mpc^{-1}$, $\Omega_0 = 0.3$, $\Omega_\Lambda
= 0.7$. The virial quantities are computed at radius
$r_{200}$\footnote{$r_\Delta$ is the radius within which the mean
  density is $\Delta$ times the critical density of the Universe.} .

\section{The data}
\label{sec: data}
The photometric information has been obtained from the SDSS DR7
database\footnote{http://cas.sdss.org/astro/en/tools/chart/chart.asp}
after searching for the galaxies that have $238\fdg983 < \rm{RA} <
240\fdg183$, $ 26\fdg633 < \rm{DEC} < 27\fdg834 $ and
$\rm{petroMag_{r'}} < 22$. The spectroscopic information has been
provided by \cite{owers2011}. The full sample is composed of 1631
galaxies with both photometric and spectroscopic information. The
cluster centre is assumed to coincide with the X-ray centre provided
by \cite{degrandi2002}.
 
\begin{figure}
\includegraphics[width=\columnwidth]{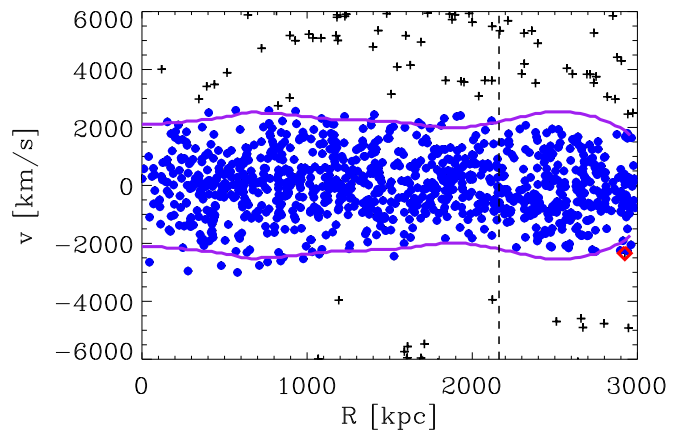}
\caption{\label{fig: rv} Distribution of the galaxies of Abell 2142 in
  the projected phase-space of projected radii and \emph{los}
  rest-frame velocities. Cluster members, as identified by both dHK
  and \emph{clean} algorithms, are denoted by blue filled dots. The
  red diamond is the galaxy identified as member by dHK but not by the
  \emph{clean} algorithm. The purple solid lines are the caustic,
  described in Sect.~\ref{sec:techniques}. The vertical dashed line
  locates the virial radius of the \concordance model (see
  Sect.~\ref{sec:mass profiles}).  }
\end{figure}

Two algorithms have been used to select cluster members, those of
\cite{denhartog1996} and of \cite*{mamon2013}, hereafter dHK and
\emph{clean}, respectively. Both identify cluster members on the basis
of their location in projected phase-space\footnote{$R$ is the
  projected radial distance from the cluster centre. We assume
  spherical symmetry in the dynamical analyses. The rest-frame
  velocity is defined as $v =
  c\,\left(z-\overline{z}\right)/\left(1+\overline{z}\right)$.  The
  mean cluster redshift $\overline{z}$ is re defined at each new
  iteration of the membership selection, until convergence.}: $R$,
$v_{\rm rest}$, using the spectroscopic values for the velocities. We
adopt the membership determination of dHK, resulting in 996
members. In fact, the \emph{clean} algorithm removes one more galaxy
that is very close to the distribution of selected members and
therefore seems unlikely to be an interloper. Anyway, this galaxy is
$\approx 3\,\Mpc$ from the cluster centre, which should make no
difference in the analysis here. Fig. \ref{fig: rv} shows the location
of galaxies in projected phase-space with the identification of
cluster member galaxies using the two methods. We use the method
described in Appendix~B of \cite{mamon2013} to obtain a preliminary
estimate of the virial radius from the velocity dispersion of the
cluster members. The value we obtain is $2.33\, \Mpc$. This is used
later as an initial--guess value for the virial radius, only to be
successively refined with more sophisticated techniques (see
Sect. \ref{sec: \concordance}).

The cluster mean redshift and \emph{los} velocity dispersion, as well
as their uncertainties, have been computed using the biweight
estimator \citep{beers1990} on the redshifts and rest--frame
velocities of the members: $\langle z \rangle = 0.08999 \pm 0.00013$,
$\sigma_{\rm los} = 1193^{+58}_{-61} \, \km/s$

\subsection{The colour identification}
\begin{figure}
\includegraphics[width=\columnwidth]{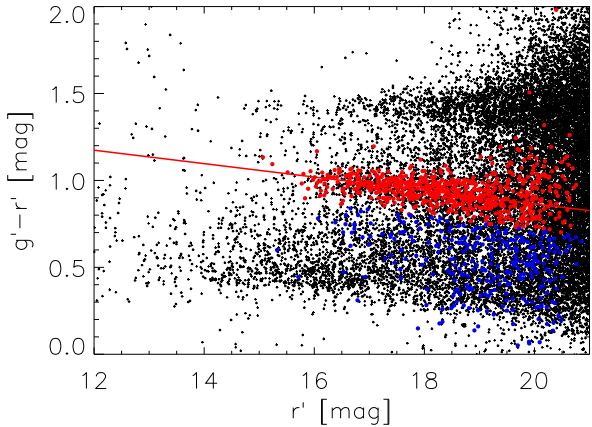}
\caption{\label{fig: cmr} Colour magnitude diagram $g'-r'$
  vs. $r'$. Red (blue) points are relative to red (blue) member
  galaxies. Black points are galaxies for which we have photometric
  information, that are not identified as members. The red solid line
  locates the red sequence.}
\end{figure}

We identify the red sequence iteratively by fitting the $g'-r'$
vs. $r'$ colour-magnitude relation of galaxies with $r'<19.5$ and
$g'-r'>0.7$, then selecting galaxies within $\pm 2\,\sigma$ of the found
sequence (where $\sigma$ is the dispersion around the best fit
relation). We refer to the cluster members within $\pm 2\,\sigma$ of the
red sequence, and those above this range, as red sequence galaxies,
and to the cluster members more than $2\,\sigma$ below the red sequence
as blue galaxies, as shown in Fig. \ref{fig: cmr}.

\subsection{Removal of substructures}
\label{sec:subs}
\cite{owers2011} found some substructures  in A2142, probably groups
that have been recently accreted by the cluster. These substructures
can alter the kinematics of the system since they still retain memory
of the infall kinematics. For this reason, we compute the mass profile
of the system excluding the galaxies belonging to these
substructures. In particular we consider the largest substructures in
this cluster, namely S2, S3, and S6, following the nomenclature of
\cite{owers2011}. Therefore, we remove galaxies inside circles, the
centres and radii of which are reported in Table \ref{Tab: nosub}.

\begin{center}
  \begin{table}
    \setlength\extrarowheight{5pt}
    \centering
    \caption{\label{Tab: nosub} Coordinates with respect to the
      cluster center, radii, and number of galaxies of the three main
      substructures, as found by \cite{owers2011}.}
    \begin{tabular}{l|c|c|c|c}
         & $x_c \, [\Mpc]$ & $y_c \, [\Mpc]$ & $r \, [\Mpc]$ & $N_{gal}$ \\
      \hline
      S2 & 0.600 & 0.763 & 0.467 & 49\\
      S3 & 2.007 & 1.567 & 0.700 & 54\\
      S6 & 2.327 & --0.180 & 0.812 & 53\\
    \end{tabular}
  \end{table}
\end{center}

\subsection{The samples}
\label{sec:samples}
Some of the techniques (described in Sect.~\ref{sec:techniques}) that
we use to compute the mass profile of the cluster rely upon the
assumption of equilibrium of the galaxy population. Red galaxies are
more likely an older cluster population than blue galaxies, probably
closer to dynamical equilibrium \citep[e.g.][]{moss1977,
  vandermarel2000}.  For this reason, red galaxies constitute a better
sample for such techniques. Among red galaxies, those outside
substructures (see Sect.~\ref{sec:subs}) are the most likely to be in
dynamical equilibrium. We therefore use these galaxies for determining
the mass profile.

The three samples that are used hereafter are as follows. We refer to
the sample made of all the member galaxies to as the ALL sample. BLUE
is the sample made of blue galaxies, and RED is the sample made of red
galaxies not belonging to the substructures described in
Sect.~\ref{sec:subs}. See Table \ref{Tab: samples} for a summary of
the number of galaxies belonging to each sample. The ALL and BLUE
samples do contain substructures.

\begin{center}
  \begin{table}
    \setlength\extrarowheight{5pt}
    \centering
    \caption{\label{Tab: samples} Number of galaxies in the three
      samples.}
    \begin{tabular}{l|c|c}
      Sample & $n_{tot}$ & $n_{200}$ \\
      \hline
      ALL  & 996 & 706 \\
      RED & 564 & 447 \\
      BLUE & 278 & 162 \\
    \end{tabular}
    \tablefoot{For each sample, the total number of member galaxies
      and the number of member galaxies within $\rtwo$ are shown, the
      latter being the value of the \concordance model (see
      Sect.~\ref{sec:mass profiles}). }
  \end{table}
\end{center}

\section{The techniques}
\label{sec:techniques}
In this section, we briefly describe the main features of the three
different techniques used in this work to compute the mass profile of
A2142. Besides the virial values of radius and mass, we obtain
estimates of the mass scale radius, which is where the logarithmic
slope of the total density profile is equal to $-2$, from which it is
possible to recover the cluster mass profile. These methods all assume
spherical symmetry.

\subsection{Methods}
\begin{description}

\item[DK: ] The dispersion kurtosis technique, hereafter shortened to
  DK, first introduced by \cite{lokas2002}, relies upon the joint fit
  of the \rm{los} velocity dispersion and kurtosis profiles of the
  cluster galaxies. In fact, fitting only the \rm{los} velocity
  dispersion profile to the theoretical relation coming from the
  projection (see \citealp{ML05b} for single integral formulae for the
  case of simple anisotropy profiles) of the \cite{Jeans1904} equation
  (see e.g. \citealp{binney1982,binney1987}) does not lift the
  intrinsic degeneracy between mass profile and anisotropy profile
  determinations (as \citealp{lokas2003} showed for the Coma cluster).
  This technique assumes dynamical equilibrium of the system, and it
  allows us to estimate the virial mass, the mass scale radius and the
  value of the cluster velocity anisotropy, considered as a constant
  with radius.\footnote{\cite{richardson_fairbairn13} have recently
    extended the DK method to more general anisotropy profiles.}

\item[MAMPOSSt: ] The MAMPOSSt technique, recently developed by
  \cite{mamon2013}, performs a maximum likelihood fit of the
  distribution of galaxies in projected phase space, assuming models
  for the mass profile, the anisotropy profile, the projected number
  density profile and the 3D velocity distribution. In particular, for
  our analysis we used different NFW models for the mass and the
  projected number density profiles, either a simplified Tiret profile
  \citep{tiret2007} or a constant value for the anisotropy profile and
  a Gaussian profile for the 3D velocity distribution. As in the DK
  method, MAMPOSSt assumes dynamical equilibrium of the system. By
  this method we estimate the virial mass, the scale radius of the
  mass density profile, and the value of anisotropy of the tracers.

\item[Caustic: ] The caustic technique, introduced by
  \cite{diaferio1997}, is different from the other two methods because
  it does not require dynamical equilibrium. As a result, this
  technique also provides the mass distribution beyond the virial
  radius. In projected phase space (see Fig. \ref{fig: rv}), member
  galaxies tend to lie in a region around $v_{\rm los} = 0\,\rm
  km\,s^{-1}$. Measuring the velocity amplitude ${\cal A}$ of the
  galaxy distribution gives information about the escape velocity of
  the system. In turn, the escape velocity is related to the
  potential, hence the mass profile: $M(r) = M(r_0) +
  (1/G)\,\int_{r_0}^r {\cal A}^2(s)\,{\cal F}_\beta(s)\,ds$, where
  ${\cal F}_\beta(r) = -2\pi G \, (3-2\beta)/(1-\beta)\,r^2
  \rho(r)/\Phi(r)$ \citep{diaferio1999}. Because ${\cal F}_\beta$ is
  usually approximated with a constant value
  \citep{diaferio1999,serra2011}, it is customary to call it a
  ``parameter''.

\end{description}

Since the DK and MAMPOSSt techniques make use of the assumption of
dynamical equilibrium of the system, the use of the RED sample allows
a more correct application of those techniques, since this sample is
likely to be the most relaxed sample.  In fact these methods just need
a tracer that obeys the Jeans equation. As long as we consider a
collisionless tracer, spherical symmetry, no streaming motions, and a
stationary system, DK and MAMPOSSt are able to reliably recover the
mass content of the cluster. On the other hand, we use the ALL sample
for the caustic technique.

As discussed in Sect. \ref{sect: intro}, some studies suggest an
elliptical morphology of the system, with evidence of some recent
intermediate mass-ratio mergers. Although this might violate the
assumptions of both spherical symmetry and equilibrium required in DK
and MAMPOSSt, and the spherical symmetry alone for the Caustic
technique, these methods are not strongly affected by this. In fact,
they have been tested on $\Lambda$CDM haloes extracted from
simulations. Although these haloes are neither spherical nor fully
relaxed, and they present substructures, the DK \citep{sanchis2004},
MAMPOSSt \citep{mamon2013} and Caustic (\citealp{serra2013};
\citealp{gifford2013}) techniques provide reliable estimates of halo
masses. As we see below (Sect. \ref{sec:mass profiles}), the fairly
close results of these dynamical methods with those from the weak
lensing analysis (which does not assume equilibrium) of
\cite{umetsu2009} suggest that this cluster cannot be far from
dynamical equilibrium.

In all three methods, we consider the scale radius of the galaxy
distribution and the scale radius of the mass distribution as two
separate and independent parameters.

\subsection{Practical implementation}
To compute the parameter values with the MAMPOSSt technique, we have
considered the galaxies of RED sample within the ``first guess''
virial radius, presented in Sect. \ref{sec: data}. As discussed in
\cite{mamon2013} (in particular see their Table 2), MAMPOSSt does not
critically depend on this choice.\footnote{However, beyond $\approx
  2.5\,r_{200}$, the infall streaming motions are important enough
  that the usual Jeans equation is inadequate for determining the
  radial velocity dispersion \citep{falco2013}.} We then performed a
Markov Chain Monte Carlo (MCMC) procedure (see e.g.  \citealp{LB02}),
using the public CosmoMC code of
A.~Lewis.\footnote{http://cosmologist.info/cosmomc}
In MCMC, the parameter space is sampled following a procedure that
compares the posterior (likelihood times prior) of a point in this
space with that of the previous point, and decides whether to accept
the new point following a criterion that depends on the two
posteriors. We use the Metropolis-Hastings algorithm. The next point
is chosen at random from a hyperellipsoidal Gaussian distribution
centred on the current point.  This procedure ensures that the final
density of points in the parameter space is proportional to the
posterior probability.  MCMC then returns probability distributions as
a function of a single parameter, or for several parameters together.
Here, the errors on a single parameter are computed by marginalizing
the posterior probabilities over the other two free parameters.

For the caustic technique, we use the ALL sample, since the
equilibrium of the sample is not required, also considering the
galaxies beyond the virial radius. To apply the caustic technique, the
${\cal F}_\beta$ parameter \citep{diaferio1999} must be chosen. The
choice of the parameter is quite arbitrary, so we tested three
different choices: the constant value 0.5, as first suggested in
\cite{diaferio1999}; the constant value 0.7 as suggested in
\cite{serra2011}; and the profile described in \cite{biviano2003}.
\begin{figure}
\includegraphics[width=\columnwidth]{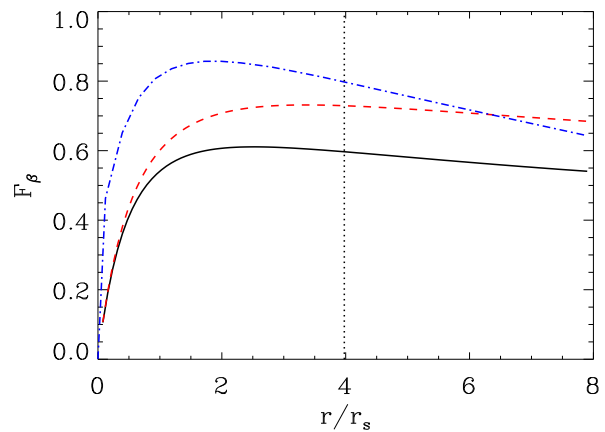}
\caption{\label{fig: fbeta} ${\cal F}_\beta$ parameter as a function
  of clustercentric distance for an NFW model. Black solid line refers
  to the isotropic case, while red dashed line refers to an ML
  anisotropy \citep{ML05b} with $r_{anis} = r_s$. Blue dash dotted
  line refers to the ${\cal F}_\beta$ by \cite{biviano2003}. The
  dotted vertical line locates the virial radius of the \concordance
  model (see Sect.~\ref{sec:mass profiles}).}
\end{figure}
The last is a smooth approximation of the ${\cal F}_\beta(r)$ derived
from numerical simulations by \cite{diaferio1999}. The actual values
of ${\cal F}_\beta$ are not likely to be very different from these we
decided to test. In fact, Fig. \ref{fig: fbeta} shows that an NFW
model leads to ${\cal F}_\beta = 0.6$ at $r=4 \, r_s \simeq r_{200}$
for isotropic orbits, while for orbits with ML \citep{ML05a}
anisotropy, it produces ${\cal F}_\beta = 0.7$ at $r = 4 \,r_s \simeq
r_{200}$. As a comparison, in Fig. \ref{fig: fbeta} the profile by
\cite{biviano2003} is shown. It has higher values in the centre, but
rapidly falls in the outer regions. The value 0.5 allows us to take
both the innermost region, where the values of ${\cal F}_\beta$ are
very low, and the outer part, where the values are larger and closer
to 0.7 into account.

When using ${\cal F}_\beta= 0.7$ and the anisotropy profile of
\cite{biviano2003}, the estimated virial masses are much greater than
those obtained with the other techniques that rely on the dynamics of
galaxies, as well as the results coming from the X-ray and the weak
lensing analysis (see below). Therefore we decided to consider only
the caustic technique with ${\cal F}_\beta = 0.5$ \citep[the same
  value has been recently adopted by][]{geller2013}. Given that for
$r_s < r < 4 r_s \approx r_{200}$ one can approximate ${\cal F}_\beta
\simeq \rm cst$ typically to $\pm11\%$ accuracy, the mass profile
returned by the caustic method changes normalization but not the shape
for different values of ${\cal F}_\beta $. Therefore this method turns
out to be very useful for constraining the mass profile shape, since
it does not assume a parametric profile like an NFW, so that we can
check whether the assumption of NFW for the mass profile is a good
one.  We adopt $r_0=0$, which relieves us from the choice of a mass at
some finite radius $r_0$.  Once we have computed the mass profile, we
fit it with an NFW profile to obtain an estimate of the mass scale
radius.

\subsection{The scale radius of galaxy distribution}
The NFW scale radius of the galaxy distribution is used as input for
the DK and MAMPOSSt analyses, therefore it has been computed for the
RED sample. The number density profile of the spectroscopic sample is
affected by the incompleteness issue.  We need to known the
distribution of tracers along the \emph{los}. Assuming spherical
symmetry, we can adopt the deprojection of the tracer surface density
profile, but we must first correct for spectroscopic
incompleteness. \citeauthor{owers2011} (2011, see their Fig. 2) have
measured their spectroscopic incompleteness in various magnitude
bins. Since their incompleteness depends rather little on magnitude,
we adopt their cumulative incompleteness measured for $R \leq
20.5$. This completeness has then been corrected in order to take into
account the artificial reduction of the number of galaxies due to the
presence of a bright star in the cluster field. Also, since we do not
wish to consider galaxies inside substructures, we also have to
correct the completeness to account for the removal of the
substructures. We divided the cluster in radial bins and counted the
galaxies inside each bin. In the bins where the presence of the star
and the removal of substructures causes a lack of detection, the area
of the bin is artificially reduced, and the mean density of galaxies
is computed in the remainder of the bin. This value is then assigned
to the whole bin.

\begin{figure}
\includegraphics[width=\columnwidth]{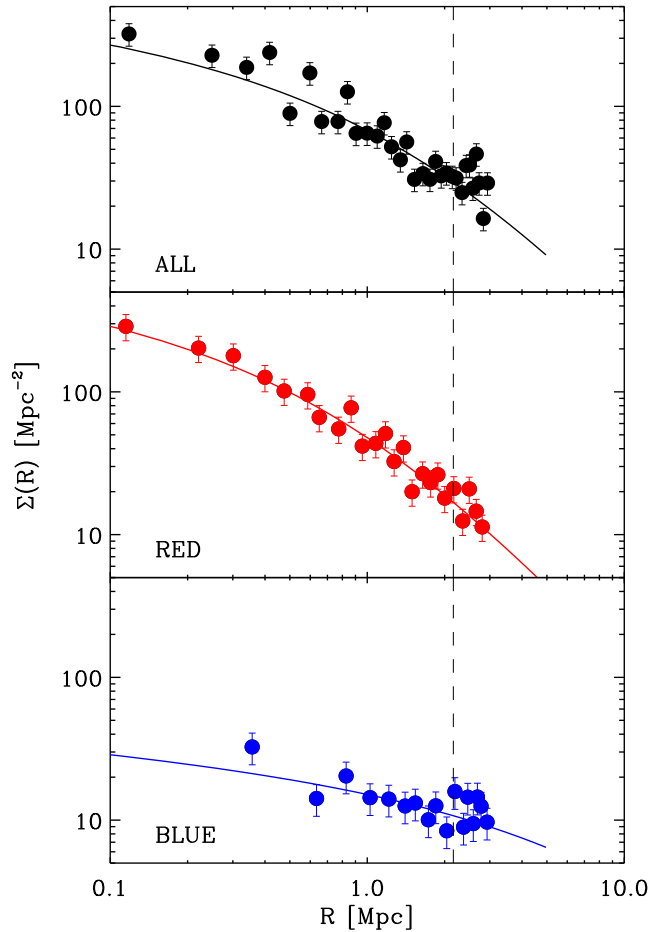}
\caption{\label{fig: ndp} Surface number density profiles for the ALL,
  RED, and BLUE samples, along with their best-fit projected NFW
  profiles. The dashed vertical line locates the virial radius of the
  \concordance model (see Sect.~\ref{sec:mass profiles}).}
\end{figure}

The RED galaxy number density profile is well fitted by a projected
NFW profile \citep{bartelmann1996,lokas2001}. The fit is an MLE fit
performed on all RED members that provide a scale radius $0.95 \pm
0.14\, \Mpc$. The ALL and BLUE samples are less concentrated, the
values of the scale radius being $1.84 \pm 0.25 \, \Mpc$ for the ALL
sample and $16 \pm 11\, \Mpc$ for the BLUE sample. A KS test (e.g.
\citealp{press1993}) provides an estimate of the reliability of these
fits. The probabilities of obtaining greater discrepancy by chance for
the RED and BLUE samples are $P = 0.95$ and $0.20$, respectively,
indicating that the model adequately fits the data. However, for the
ALL sample, the corresponding probability is only $P = 0.05$,
indicating that the model only marginally fits the data.  In
Fig. \ref{fig: ndp} the surface number density profiles for the
different samples are shown. The scale radius for the BLUE sample is
very high and is due to a very flat distribution of these galaxies.

\section{Mass profiles}
\label{sec:mass profiles}

\subsection{Mass profiles obtained from the different methods}

\begin{figure}
\includegraphics[width=\columnwidth]{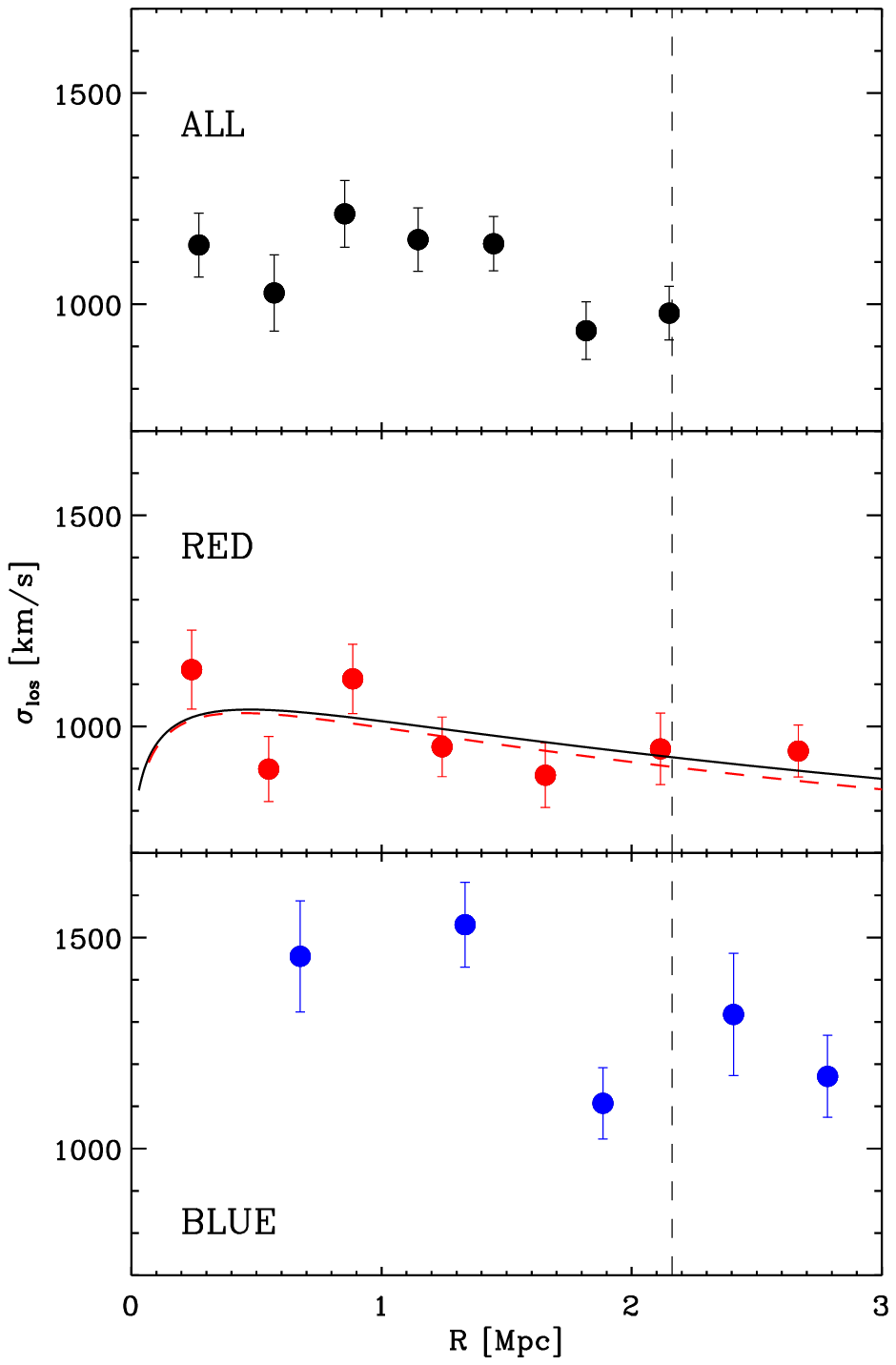}
\caption{\label{fig: vdp} Velocity dispersion profiles for the ALL,
  RED, and BLUE samples. For the RED sample we also show the
  best-fit profile coming from the DK analysis (black), and the
  profile computed after the MAMPOSSt analysis (dashed red). The
  dashed vertical line locates the virial radius of the \concordance
  model (see Sect.~\ref{sec:mass profiles}).}
\end{figure}

We used the velocities of the galaxies within the ``first guess''
virial radius (see Sect. \ref{sec: data}) to compute the mass profile
of A2142. In Figure \ref{fig: vdp}, the velocity dispersion profiles are
shown, along with the best-fit profiles coming from the DK and
MAMPOSSt analyses.

The DK technique assumes a constant value for the anisotropy, while we
have chosen two profiles for the anisotropy model in MAMPOSSt, a
constant value and a Tiret profile $\beta(r) = \beta_0 + (\beta_\infty
- \beta_0)\, r / (r + r_{\rm anis})$. Here, we set $\beta_0 = 0$
(inner isotropy) and set $r_{\rm anis}$ to the scale radius of the
galaxy's number density profile. The maximum values of the likelihoods
are similar when using the two anisotropy models, therefore for the
sake of simplicity we consider only the case of a constant velocity
anisotropy. In Sect.~\ref{sec:anisotropy profile}, we compute the
anisotropy profile for the RED sample and find that indeed it is
compatible with a constant value.

We also tried to assume different mass profiles and velocity
anisotropy models in MAMPOSSt, namely a Burkert \citep{burkert1995}, a
Hernquist \citep{hernquist1990} and a softened isothermal sphere
profile (e.g. \citealp{mamon87}; \citealp{geller1999}), all with both
constant and Tiret anisotropy profiles.  However our data-set is not
large enough to allow us to distinguish between these different
models. All provide acceptable fits. As a consequence, the resulting
estimates of virial mass and mass profile concentration are very
similar to the case of NFW mass profile with constant anisotropy, with
differences of very few percent. We therefore only considered the NFW
model for the mass profile.

The results are summarized in Table~\ref{Tab: all}.  Figure~\ref{fig:
  mamposst params} shows the detailed results of our MAMPOSSt MCMC
analysis. The mass scale radius is not well constrained by
MAMPOSSt. This does not affect the subsequent analysis, since in
Sect. \ref{sec: \concordance}, we perform a weighted mean of the
results from the different methods.

In Fig. \ref{fig: mass all}, we show the mass profiles obtained from
the different methods, along with the virial values of mass and
radius. The results coming from the X-ray \citep{akamatsu2011} and
weak lensing \citep[][WL hereafter]{umetsu2009} analysis are also
shown.

\begin{figure*}
\centering
\includegraphics[width=0.8\textwidth]{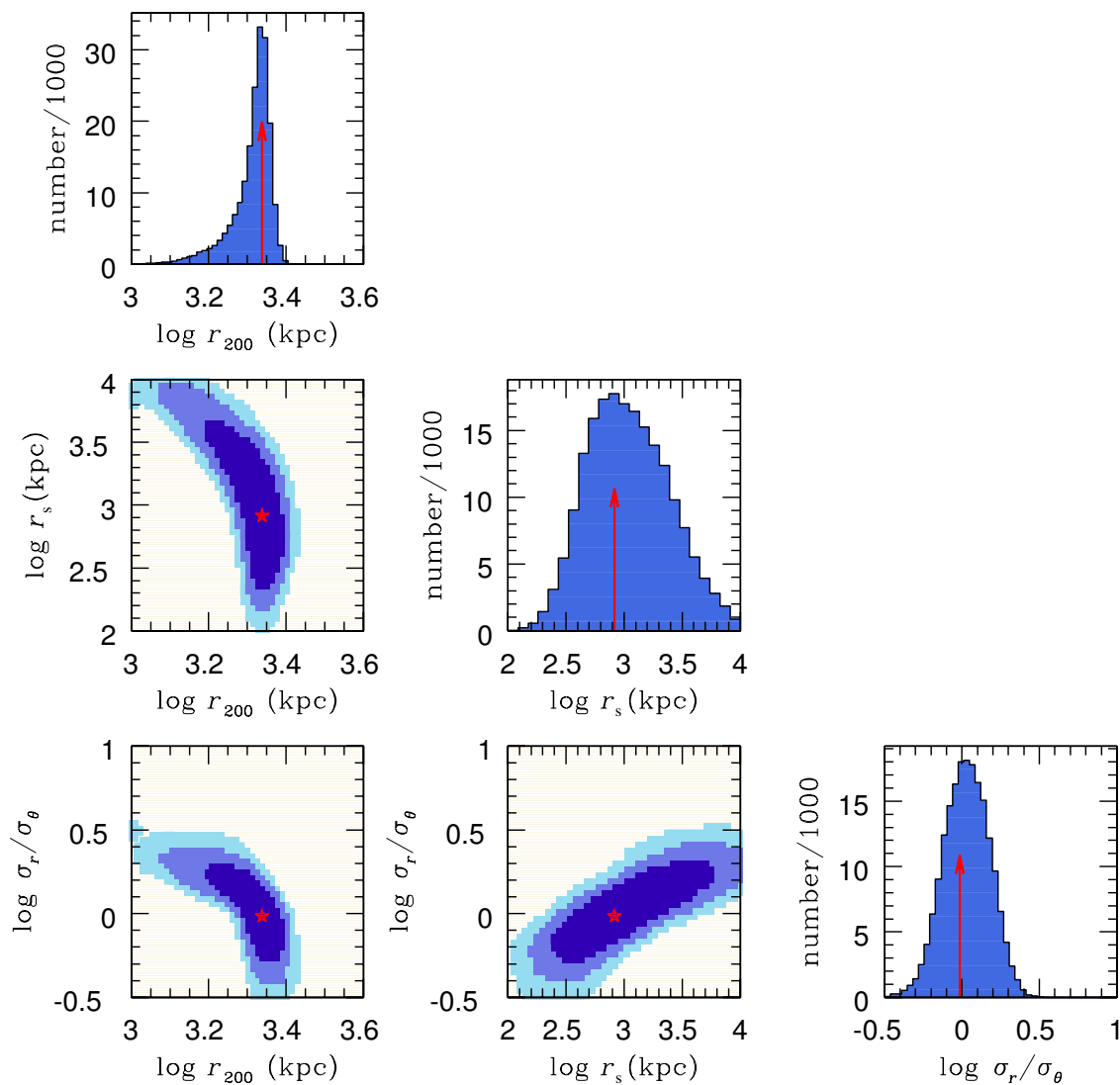}
\caption{\label{fig: mamposst params} Parameter space and probability
  distribution functions for the virial radius, mass profile scale
  radius, and velocity anisotropy, as found by MAMPOSSt. The coloured
  regions are the 1,2,3 $\sigma$ confidence regions, while the red
  stars and the red arrows locate the best-fit values. These are based
  upon an MCMC analysis with 6 chains of 40$\,$000 elements each, with
  the first 5000 elements of each chain removed (this is the
  \emph{burn-in} phase that is sensitive to the starting point of the
  chain). The priors were flat within the range of each panel, and
  zero elsewhere.}
\end{figure*}
\begin{center}
  \begin{table*}
    \setlength\extrarowheight{5pt}
    \centering
    \caption{\label{Tab: all} Virial quantities of Abell 2142 obtained
      from different techniques}
    \begin{tabular}{lcllllcl}
      \hline
      \hline
Method & Sample      & \multicolumn{1}{c}{$M_{200} \, [10^{15} \Msun]$} &
$r_{200} \, [\Mpc]$ & $r_s \, [\Mpc]$ & \multicolumn{1}{c}{$c$} &
$\sigma_r/\sigma_t$ & \multicolumn{1}{c}{$\beta$}\\
      \hline
      Caustic (${\cal F}_\beta=0.5$) & ALL & $1.26^{+0.54}_{-0.42}$ &
      $2.17^{+0.27}_{-0.28}$ & $0.58^{+0.12}_{-0.10}$ & $3.7$$\pm$$0.9$ & & \\
      DK & RED & $1.32^{+0.11}_{-0.21}$ & $2.20^{+0.06}_{-0.12}$ &
      $0.93^{+0.39}_{-0.10}$ & $2.4$$\pm$$0.6$ & $1.0^{+0.20}_{-0.04}$ & $0.0_{-0.1}^{+0.3}$\\
      MAMPOSSt & RED & $1.28^{+0.14}_{-0.49}$ & $2.18^{+0.08}_{-0.32}$ & 
      $0.83^{+1.73}_{-0.35}$ & $2.6^{+2.0}_{-1.9}$ &  $1.0^{+0.50}_{-0.20}$ &
          $0.0$$\pm$$0.6$ \\
      \hline
      Kinematics & & $1.31^{+0.26}_{-0.23}$ & $2.19$$\pm$$0.14$ & $0.64$$\pm$$0.17$ & $3.4$$\pm$$0.9$ & \\ 
      \hline
      X-ray & & $1.11^{+0.55}_{-0.31}$ & $2.08^{+0.30}_{-0.22}$ & $0.74$$\pm$$0.31$ & $2.8$$\pm$$1.1$ & \\
      WL & & $1.24^{+0.18}_{-0.16}$ & $2.16$$\pm$$0.10$ & $0.51$$\pm$$0.08$ & $4.3$$\pm$$0.7$ & \\
      \hline \hline
      Combined model & & $1.25$$\pm$$0.13$ & $2.16$$\pm$$0.08$ & $0.54$$\pm$$0.07$ & $4.0$$\pm$$0.5$ & \\
      \hline
      \hline
    \end{tabular}
    \tablefoot{Values of virial mass, virial radius, mass scale
      radius, concentration, and two measures of the velocity
      anisotropy, for different techniques. Also shown are the average
      value of the kinematical techniques after symmetrizing the
      errors and the value of the \concordance model, obtained as the
      result of the average of all the values coming from the
      different techniques (see Sect.~\ref{sec:anisotropy profile} for
      the average procedure). X-ray values come from
      \cite{akamatsu2011}, weak lensing (WL) from
      \cite{umetsu2009}. Both for X-ray and WL we had the values and
      the errors of the virial radius and the concentration: we have
      symmetrized these errors and propagated them to obtain the
      estimates of the errors on the mass scale radii.}
  \end{table*}
\end{center}

\begin{figure}
\includegraphics[width=\columnwidth]{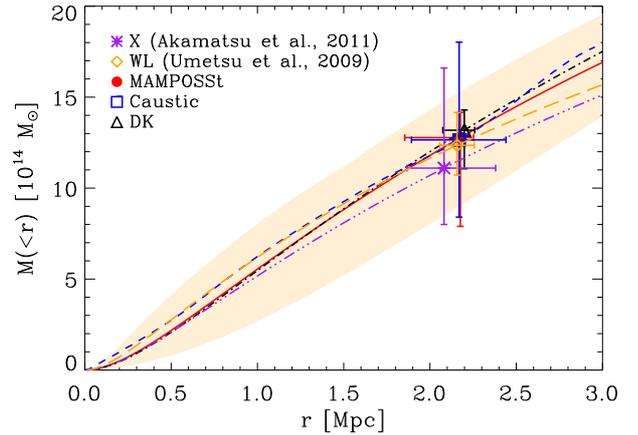}
\caption{\label{fig: mass all} Mass profiles computed from the
  different methods.  The black dash-dotted line and the triangle with
  error bars refer to DK technique, the dashed blue line and blue
  square to the caustic method, the solid red line and red point to
  MAMPOSSt. The symbols with error bars refer to the virial mass and
  radius. The purple asterisk with error bars and the purple dash
  triple dotted line are the result of the X-ray analysis, while the
  orange diamond with a long dashed line is the one coming from weak
  lensing analysis. The shaded area is the $1\,\sigma$ confidence
  region of the mass profile according to the MAMPOSSt results.}
\end{figure}

\subsection{Combined mass profile}
\label{sec: \concordance}
We combined the constraints from the different mass modelling methods
to build a \emph{\concordance} mass profile, assuming again an NFW
density profile. We attempted to give the same weight to kinematics,
X-ray, and WL in the final estimate of the parameters, so we computed
single values coming from kinematical techniques for the mass scale
radius and virial radius. For this, we took the mean of the values
$r_s$ and $r_{200}$ of the different methods, inversely weighting by
the symmetrized errors. Since the measures of these two quantities by
the various methods are not independent (as they are based on
essentially the same data sets), we multiplied the error on the
average by $\sqrt{3}$, 3 being the number of values used to compute
the average. In fact, the usual error on the weighted average
decreases like the square root of the number of values.

\begin{figure}
 \includegraphics[width=\columnwidth]{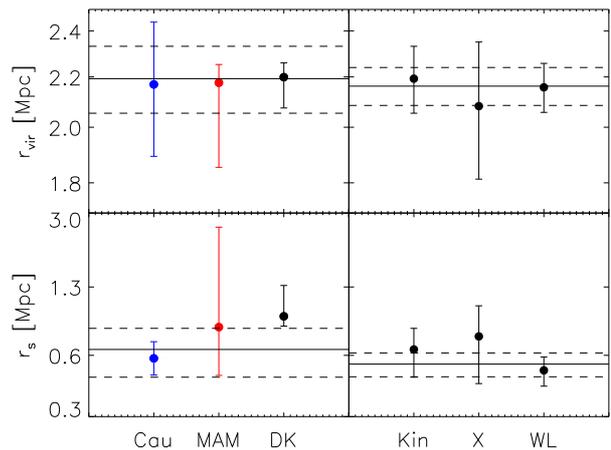}
\caption{\label{fig: param comparison} Virial (top panels) and mass
  scale (bottom panels) radius for all the methods. \emph{Left
    panels:} blue diamonds are values obtained from the caustic
  technique, red ones for MAMPOSSt, and black ones for DK (from left
  to right, respectively). The average value and its error are the
  solid and dashed lines, respectively. See the text for the
  computation of the error. \emph{Right panels:} values obtained from
  the kinematical analysis, X-ray and WL (from left to right,
  respectively). The average value and its error are the solid and
  dashed lines, respectively.}
\end{figure}
The mean value and its error are shown in the left panels of
Fig. \ref{fig: param comparison}. In the right panels of
Fig. \ref{fig: param comparison}, we plot the values of mass scale and
virial radius obtained from the three independent methods: kinematics,
X-ray, and WL. The average error-weighted value and its error, this
time computed without multiplication factor (since the three measures
are independent), are $r_{200}=2.16 \pm 0.08\,\rm Mpc$, $r_s=0.54 \pm
0.07\,\rm Mpc$.

\section{Velocity anisotropy profiles}
\label{sec:anisotropy profile}
Although with DK and MAMPOSSt we have assumed some models for the
velocity anisotropy profile, we now wish to determine it in a
non-parametric way using the Jeans equation. For this, we use the mass
profile we obtained by combining the information coming from the three
dynamical methods, X-ray and WL. The Jeans equation contains four
unknown quantities, therefore to solve it we need three other
relations, namely the Abell integrals to relate the projected number
density and velocity dispersion to the real ones and assume a mass
profile for the cluster. This \emph{anisotropy inversion} was first
solved by \cite{binney1982}, but several other authors have provided
simpler algorithms.  We follow the approach of \cite{solanes1990}, and
we tested the results by comparing them with those obtained following
the approach of \cite{dejonghe1992}. Once the mass profile is
specified, this procedure is fully non--parametric. In fact, instead
of fitting the number density profile, we binned and smoothed it with
the LOWESS technique \citep[see e.g.][]{gebhardt1994}. We then
obtained the 3D number density profile by using Abel's equation (e.g.
\citealp{binney1982}). In the same way, we smoothed the binned
$\sigma_{\rm los}$ profile. This procedure requires the solution of
integrals up to infinity. \cite{mamon2010} show that a $3\,\sigma$
clipping removes all the interlopers beyond 19 virial
radii. Therefore, an extrapolation up to such a distance is enough to
solve the integrals having infinity as the integration limit. We used
30 \Mpc\ as the maximum radius of integration, and extrapolate the
smoothed profiles up to this limit. A factor--2 change in the upper
limit of integration does not affect our results in a significant way.

The result of the anisotropy inversion is shown in Fig. \ref{fig: srst
  comparison}. The confidence levels were obtained by estimating two
error contributions. One contribution comes from the uncertainties in
the surface density and $\sigma_{\rm los}$ profiles. Since 80\% of the
relative uncertainty of the product $\Sigma \sigma_{\rm los}^2$ comes
from the uncertainty of $\sigma_{\rm los}$ \citep{trilling2014}, we
only considered the error contribution from the latter. It is
virtually impossible to propagate the errors on the observed
$\sigma_{\rm los}$ through the Jeans inversion equations to infer the
uncertainties on the $\beta$ profile solution. We then proceeded to
estimate these uncertainties the other way around. We modify the
$\beta$ profile in two different ways: 1) $\beta(r) \rightarrow
\beta(r) + S + T\, r$, and 2) $\beta(r) \rightarrow J \, \beta(r) +
Y$, using a wide grid of values for the constants, respectively $(S,
T)$ and $(J, Y)$. Using the mass and anisotropy profiles, it is then
possible to determine $\sigma_r(r)$ and then the $\sigma_{\rm los}$
profile \citep[e.g.][]{ML05b}. The range of acceptable $\beta$
profiles is determined by a $\chi^2$ comparison of the resulting
$\sigma_{\rm los}$ profiles with the observed one.

In addition, another source of uncertainty on the $\beta$ profile
solution comes from the uncertainty in the mass profile. This is
estimated by running the anisotropy inversion for four different mass
profiles corresponding to the combination of allowed values of virial
and mass scale radii within $1\,\sigma$. The profiles obtained
modifying the mass profile (not shown) lie within the confidence
interval of the main result, so that the confidence interval
represents the uncertainty on the anisotropy profile well.

\begin{figure}
\includegraphics[width=\columnwidth]{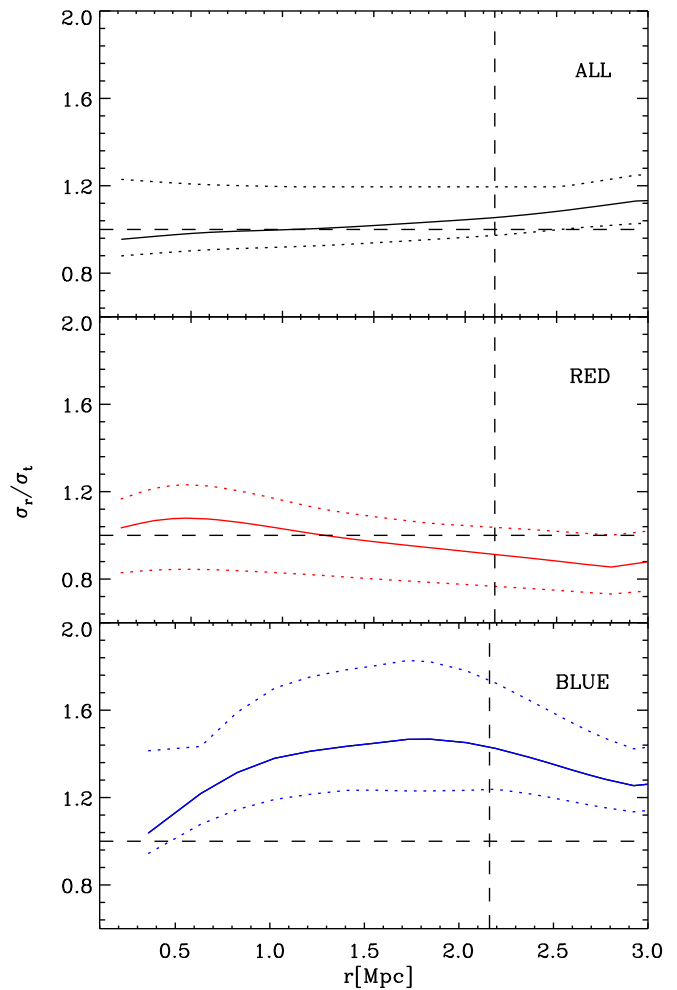}
\caption{\label{fig: srst comparison} Velocity anisotropy profiles for
  the ALL, RED, and BLUE samples. The solid line is the result of the
  inversion of the Jeans equation, while the dotted lines are the
  $1\,\sigma$ confidence intervals. The vertical dashed line locates
  the virial radius.}
\end{figure}

The ALL sample $\beta(r)$ depends weakly on radius: the innermost
region is compatible with isotropy, while the anisotropy is
increasingly radial at large radii. The RED sample is compatible with
isotropy at all radii. The difference between the two samples is
almost entirely due to the BLUE galaxies, the anisotropy of which is
compatible with isotropy in the centre, then becomes rapidly radially
anisotropic, and finally flattens at radii $>1\Mpc$.

As a check, we compare the values of $\beta$ obtained from the
anisotropy inversion with the best-fit results of DK and MAMPOSSt. In
these techniques, we assumed a constant value of the anisotropy for
the RED sample, which appears to be a good assumption given the
results of $\beta$ after the inversion. The value estimated by both DK
and MAMPOSSt is $\beta = 0.0$, consistent within the uncertainties
with the $\beta$ profile shown in Fig. \ref{fig: srst comparison}.

\section{$Q(r)$ and $\beta - \gamma$ relations}
\label{PPSD}

Since our anisotropy inversion provides us with the radial variations
of $\sigma(r)=\sqrt{\sigma_r^2+2\sigma_t^2}$, $\beta(r)$ (from which
$\sigma_r(r)$ follows), we can take advantage of the results just
found for the galaxy populations of A2142 to test the PPSD profile and
the relation linking the logarithmic slope of the density profile and
the anisotropy $\beta(r)$.

Both the PPSD and $\beta-\gamma$ relations were derived from
dissipationless single-component simulations. It is therefore not
clear whether the power-law PPSD and the linear beta-gamma relation,
both found for the particles of single component dark matter (DM)-only
simulations, will be obtained when using galaxies to measure the
velocity dispersion or velocity anisotropy and whether one should use
the total density or the galaxy number density in these two
relations. We discuss this further in Section~\ref{discussion}.

\subsection{Use of the total matter density profile}

We begin by adopting the total density profile $\rho(r)$.
We compute both the PPSD profile $Q(r) = \rho / \sigma^3$ and its
radial counterpart $Q_r(r) = \rho / \sigma_r^3$. 
\begin{figure}
\includegraphics[width=\columnwidth]{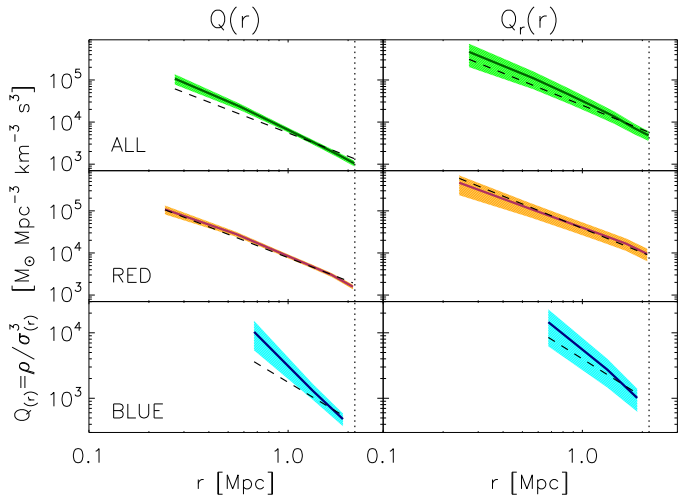}
\caption{\label{fig: qr} Radial profiles of $Q$ (left columns) and
  $Q_r$ (right columns) within the virial radius, and the $1\,\sigma$
  confidence regions (shaded areas), for different types of member
  tracers: green for the ALL sample (top panels), red for the RED
  sample (middle panels), and blue for the BLUE sample. The shaded
  areas represent the propagation of the errors associated with
  $\rho$, $\sigma$ and $\sigma_r$.  The dashed lines are the power-law
  relations $Q(r) \propto r^{-1.84}$ and $Q_r (r) \propto r^{-1.92}$
  found by \cite{dehnen2005} on numerically simulated haloes. The
  vertical dotted lines locate the virial radius of the \concordance
  model (see Sect.~\ref{sec:mass profiles}).}
\end{figure}
In Fig. \ref{fig: qr}, we show, for the different tracers (ALL, RED,
BLUE), the radial profile of $Q(r)$ (left panels) and $Q_r(r)$ (right
panels) within the virial radius. To compute the errors on the
best-fit slope parameters, we have assumed that the number of
independent $Q$ and $Q_r$ values are the same as those of the observed
velocity dispersion profile (see Fig. \ref{fig: vdp}).

  \begin{table*}
    \setlength\extrarowheight{5pt}
    \centering
    \caption{\label{Tab: qr} Best-fit parameters of the PPSD profile}
    \begin{tabular}{c|cc|cc}
    \multicolumn{1}{c}{}  & \multicolumn{2}{c}{$Q(r)$} & \multicolumn{2}{c}{$Q_r(r)$} \\
      \hline
      \hline
      & $A$ & $B$ & $A$ & $B$ \\
      & $[\Msun \; \Mpc^{-3} \; \km^{-3} \; s^3]$ &  & $[\Msun \; \Mpc^{-3} \; \km^{-3} \; s^3]$ &  \\
      \hline
      \multicolumn{5}{l}{Fixed slope}\\
      \hline
      ALL & $5534 \pm 314$ & $-1.84$ & $25071 \pm 3341$ & $-1.92$ \\
      RED & $7727 \pm 391$ & $-1.84$ & $38484 \pm 5622$ & $-1.92$ \\
      BLUE & $1753 \pm 294$ & $-1.84$ & $3998 \pm 1084$ & $-1.92$ \\
      \hline
      \multicolumn{5}{l}{Free slope}\\
      \hline
      ALL & $6342 \pm 367$ & $-2.28 \pm 0.11$ & $29175 \pm 4223$ & $-2.27 \pm 0.24$ \\
      RED & $8034 \pm 411$ & $-2.00 \pm 0.09$ & $38881 \pm 5665$ & $-1.77 \pm 0.23$ \\
      BLUE & $3121 \pm 793$ & $-2.97 \pm 0.50$ & $5413 \pm 1810$ & $-2.60 \pm 0.67$ \\
      \hline
      \hline
    \multicolumn{5}{c}{}  \\
    \multicolumn{1}{c}{}  & \multicolumn{2}{c}{$Q(r) \; GAL$} & \multicolumn{2}{c}{$Q_r(r) \; GAL$}\\
      \hline
      \hline
      & $A$ & $B$ & $A$ & $B$ \\
      & $[10^{-9} \; \Mpc^{-3} \; \km^{-3} \; s^3]$ &  & $[10^{-9} \Mpc^{-3} \; \km^{-3} \; s^3]$ & \\
      \hline
      \multicolumn{5}{l}{Fixed slope}\\
      \hline

      ALL & $    3.7 \pm    0.18$ & $-1.84$ & $    17. \pm     2.2$ & $-1.92$\\
      RED & $    2.9 \pm    0.14$ & $-1.84$ & $    13. \pm     1.9$ & $-1.92$\\
      BLUE & $   0.37 \pm   0.056$ & $-1.84$ & $   0.68 \pm    0.18$ & $-1.92$\\
      \hline
      \multicolumn{5}{l}{Free slope}\\
      \hline
      ALL & $    3.6 \pm    0.86$ & $  -1.72 \pm    0.10$ & $    17. \pm     2.4$        &       $  -1.72 \pm    0.23$ \\
      RED & $    2.9 \pm    0.59$    &     $  -1.75 \pm    0.09$ & $    14. \pm     2.0$    &     $  -1.52 \pm    0.23$ \\
      BLUE & $   0.36 \pm    0.39$    &     $  -1.74 \pm    0.48$ & $   0.62 \pm    0.21$    &     $  -1.39 \pm    0.66$ \\
      \hline
      \hline

    \end{tabular}
    \tablefoot{\erratum{The PPSD profile is parametrized as $Q(r) = A \,
      r^B$. The first panel at the top shows the results of the fit of
      $Q(r)$ and $Q_r(r)$ for the different samples, both when keeping
      the exponent fixed to the values suggested by \cite{dehnen2005},
      and when considering the exponent as a free parameter. In the
      bottom panel (identified by $Q(r) \; GAL$ and
      $Q_r(r) \; GAL$), the same quantities are shown, but they refer
      to the PPSD computed using the galaxy number density profile
      instead of the total matter density profile.}}
  \end{table*}

Assuming a power-law behaviour of the PPSD profile, as suggested by
\cite{dehnen2005}, we fit the profiles of both $Q(r)$ and $Q_r(r)$ in
two ways, either keeping the exponent fixed to the values found for
haloes in $\Lambda$CDM simulations by \cite{dehnen2005} or considering
it as a free parameter. In both cases the normalization is left as a
free parameter. In Table \ref{Tab: qr} the results of such fits are
shown.  The $Q(r)$ profile for the RED sample is consistent within
less than $2\,\sigma$ with the $r^{-1.84}$ relation by
\cite{dehnen2005}. The fit of the profile with a linear relation in
the log-log plane is compatible with the theoretical value $-1.84$
within $1.7\,\sigma$.  On the other hand, for the BLUE sample, the
slope of the PPSD is steeper than the theoretical expectation.

The $\sigma_r$ profile is affected by larger uncertainties than the
$\sigma$ profile, because the former combines the uncertainties from
the latter and $\beta(r)$, which are the parameters produced by the
anisotropy inversion algorithm of \cite{solanes1990}. It is therefore
not surprising that, within the uncertainties, the $Q_r$ profiles of
all three samples appear consistent with the theoretical expectation
for simulated $\Lambda$CDM haloes \citep{dehnen2005}, $Q_r \propto
r^{-1.92}$. We note, however, that the agreement is quite remarkable
(within $0.3\,\sigma$) for the RED sample.

\cite{ludlow2010} warn against fitting the pseudo phase--space density
profile outside the scale radius, because of the upturn they find in
the $Q(r)$ profile in the outer regions. However, for our three
samples, none of the $Q(r)$ and $Q_r(r)$ profiles show significant
curvature in log-log space.

\begin{figure}
\includegraphics[width=\columnwidth]{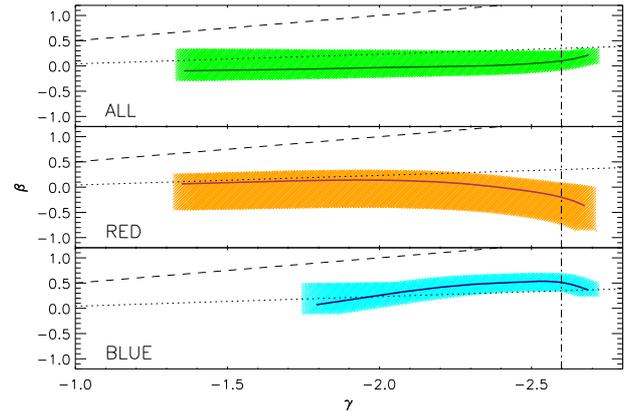}
\caption{\label{fig: betagamma} Velocity anisotropy versus logarithmic
  slope of the total density profile. The samples are ALL galaxies
  (top), RED (middle), and BLUE galaxies (bottom panel). The shaded
  areas are the $1\,\sigma$ confidence regions. The $\beta-\gamma$
  relation found by \cite{hansen2006} for single-component
  dissipationless simulations is shown as the dotted lines. The dashed
  line is the limit below which the relation by \cite{ciotti2010}
  holds. The vertical dot-dashed line locates the value of $\gamma$ at
  the virial radius.}
\end{figure}

In Fig. \ref{fig: betagamma}, we show the anisotropy - density slope
relation. The $\beta-\gamma$ relation of the ALL sample matches the
one found by \cite{hansen2006} closely in single-component
dissipationless simulations (cosmological and academic); however, the
$\beta-\gamma$ relation for the RED sample shows curvature, with lower
values of $\beta$ at the steeper slopes (larger radii) than found in
simulations by \cite{hansen2006}.

It can be proven that all multicomponent spherical systems with
positive phase-space distribution function, for which 1) the density
of a component is a separable function of total gravitational
potential and radius, 2) $\beta(0)\leq 1/2$ (i.e. are not too much
radially anisotropic at the center), necessarily satisfy $\beta(r) < -
\gamma(r)/2$, where the velocity anisotropy $\beta$ and the
logarithmic slope of density $\gamma$ are for that component, as shown
in \cite{ciotti2010} \citep[see also][]{vanhese2011}. It is not clear
whether the galaxy components of clusters of galaxies have such
separable densities.

\subsection{Use of the tracer density profile}

As we discuss at length in Sect.7, it is not obvious that one should
use the total mass density profile rather than the tracer number
density profile in evaluating the PPSD and the $\beta-\gamma$
relations, when we want to compare them to those found in numerical
simulations. As a result, we now repeat our analyses of the PPSD and
the $\beta-\gamma$ relations, replacing the total mass density with
the number density of the tracer of the sample.

\begin{figure}
\includegraphics[width=\columnwidth]{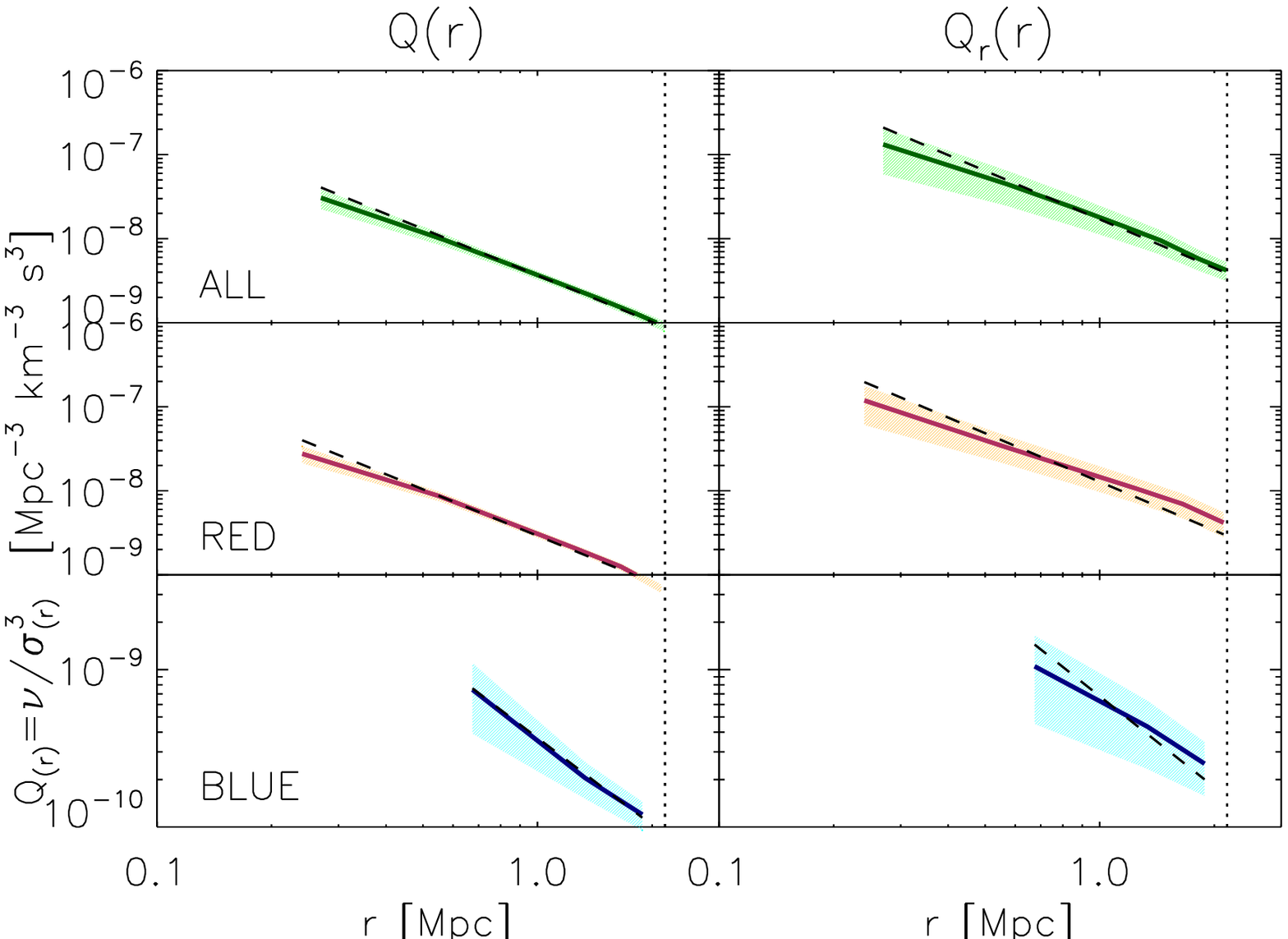}
\caption{\label{fig: qrnu} \erratum{Same as Fig. \ref{fig: qr}, but
    now using the radial profiles of galaxy number density instead of
    total mass density to estimate the PPSD.}}
\end{figure}

In Fig. \ref{fig: qrnu}, we show the PPSD computed using the galaxy
number density profile instead of the total matter density
one. \erratum{These PPSDs are either consistent with the relation of
  \citeauthor{dehnen2005} ($Q(r)$ for BLUE sample) or only slightly
  shallower, but not less consistent with that relation than found for
  the PPSDs computed with the mass density
  profile. \footnote{\erratum{Items in red in the main text are the
      consolidated version including the Corrigendum of
      \cite{munari2015}}}}

\begin{figure}
\includegraphics[width=\columnwidth]{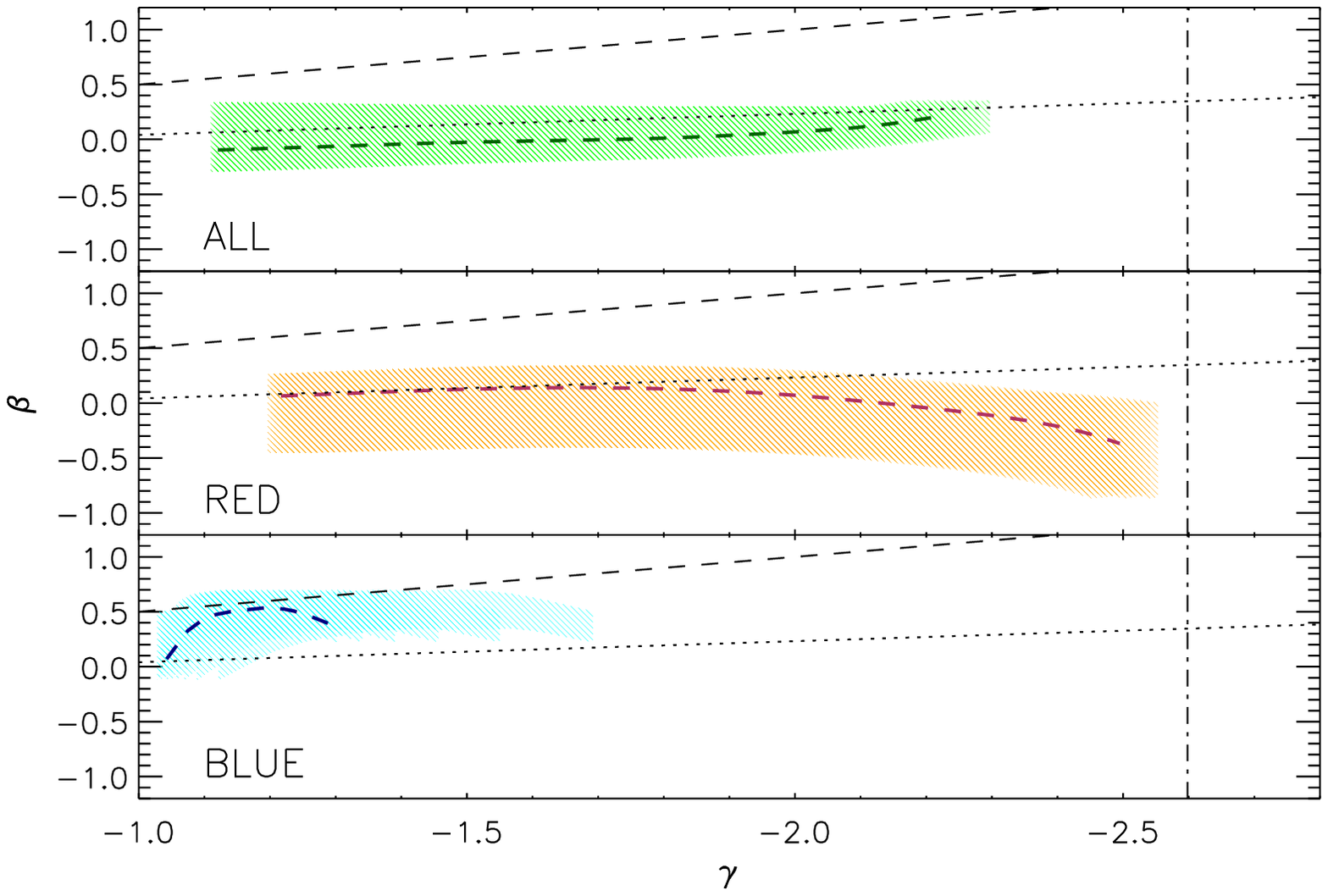}
\caption{\label{fig: betagamma gal} Same as Fig. \ref{fig: betagamma},
  but now using the radial profiles of galaxy number density of the
  three samples instead of total mass density to estimate the slope.}
\end{figure}

In Fig. \ref{fig: betagamma gal}, we show the $\beta - \gamma$
relation computed using the galaxy number density profile instead of
the total matter density one. The behaviour does not change
significantly from the case of the $\beta - \gamma$ relation computed
using the total matter density profile: the overall shapes of the
profiles are similar, but the BLUE sample now presents a noisier
profile, while ALL and RED profiles are shifted towards higher values
of $\gamma$, reflecting the shallower trend of the galaxy number
density profile with respect to the matter density one.  We discuss
these results below.

\section{Discussion}
\label{discussion}

\subsection{Dynamical status}
\cite{munari2013} report the scaling relation between the virial mass
of clusters and the velocity dispersion of the member galaxies within
the virial sphere. Using the most realistic (``AGN'') hydrodynamical
simulation at their disposal, they find
$\sigma_{1D}=1177\,[h(z)\,M_{200}/10^{15} M_\odot]^{0.364}$ for the
galaxies within the virial sphere, where $\sigma_{1D}$ is the total
$3D$ velocity dispersion within $\rtwo$, divided by $\sqrt{3}$. The
analysis was carried out in the 6D phase space, so is immune to
projection effects. The statistical nature of their relation suggests
that it should hold for real, observed, and relaxed systems. As a
test, we checked the consistency of the velocity dispersion -- mass
relation found by \cite{munari2013} with our findings for A2142. The
values of virial mass obtained with this relation are: $1.42 \times
10^{15} \Msun$ for the ALL sample, $1.07 \times 10^{15} \Msun$ for the
RED sample, and $2.50 \times 10^{15} \Msun$ for the BLUE sample. The
values obtained for the ALL and RED samples agree, within the
uncertainties, with the \concordance value of the mass of A2142. This
seems to indicate that RED cluster members are in, or very close to,
equilibrium. The large difference obtained for the BLUE cluster
members warns against using the blue galaxy \rm{los} velocity
dispersion as a proxy for the cluster mass.

A glance at Table~\ref{Tab: all} indicates that our different
estimates of the mass concentrations are bimodal: the caustic and weak
lensing have values $\simeq 4$, while those for the DK, MAMPOSSt, and
X-ray methods are $<3$. Could these lower mass concentrations found by
methods based upon internal kinematics be a sign that A2142 is out of
dynamical equilibrium?  The substructures found by \cite{owers2011}
and the results by \cite{rossetti2013} on the importance of the
mergers undergone by A2142 suggest that full relaxation is to be
excluded. On the other hand, the agreement on the virial radius
amongst the different methods and with the results from X-ray and
lensing (the latter does not require equilibrium) suggests that A2142
is not far from dynamical equilibrium.

\begin{figure}
\includegraphics[width=\columnwidth]{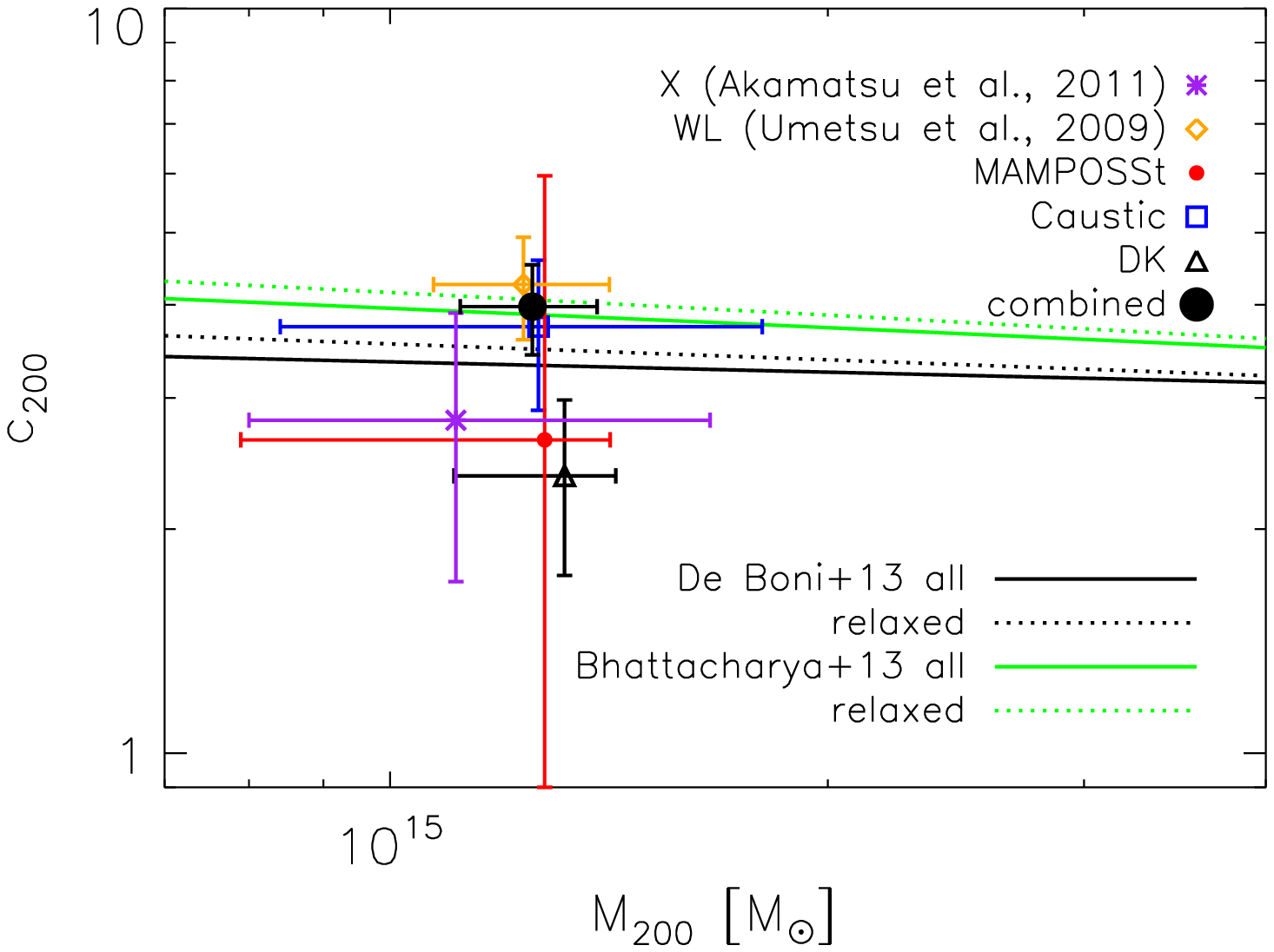}
\caption{\label{fig: c200m200} Concentration -- mass relation, with
  respect to an overdensity 200 times the critical one. Purple
  asterisk refers to the X--ray values by \cite{akamatsu2011}, orange
  diamond to the WL values by \cite{umetsu2009}, small red circle
  refers to the values obtained by MAMPOSSt, blue square by the
  caustic method, black triangle by DK, the big black circle to the
  values of the \concordance model. Lines are the theoretical
  predictions, and in black the relations by \cite{deboni2013} when
  considering all (solid) and relaxed (dotted) clusters. In green the
  relations by \cite{bhattacharya2013} when considering all (solid)
  and relaxed (dotted) clusters.}
\end{figure}

In Fig. \ref{fig: c200m200}, the concentration -- mass relation for
A2142 is shown along with theoretical relations by
\cite{bhattacharya2013} and \cite{deboni2013} based on cosmological
N--body and hydrodynamical simulations, respectively. The value of the
\concordance model [$ M_{200} = (1.25 \pm 0.13) \times 10^{15} \Msun$
  and $c = 4.0\pm 0.5$] agrees within $1\,\sigma$ with the ``relaxed''
case of \cite{deboni2013}, while it is in excellent agreement with
both the relations by \cite{bhattacharya2013}. This strengthens the
scenario of A2142 being very close to equilibrium.

\subsection{Mass density profile}
Previous studies based on the kinematics of galaxies in clusters have
shown that galaxy populations have similar concentrations to those of
the total matter, or slightly smaller, blue galaxies being instead
much less concentrated \citep[see,
  e.g.][]{biviano2003,katgert2004}. On the other hand,
\cite{biviano2009} found in the ENACS clusters that the red galaxy
population has a concentration that is as much as 1.7 times lower for
the total matter density profile. Here, we find that the scale radius
for the RED galaxy number density profile (0.95 Mpc) is 1.8 times
greater than for the total mass density profile from our \concordance
model, which agrees with the ENACS result.  \cite{collister2005}
found, on a stacked sample from the 2dFGRS, values of galaxy
concentration comparable to ours, when considering objects as massive
as A2142 (see their Fig. 7), although with uncertainties that are too
large to distinguish between red and blue samples.

The scale radius of the BLUE population in Abell 2142 appears
unusually high, leading to concentrations (using our \concordance
virial radius) of 0.16 (best) or 0.39 ($+1\,\sigma$), which are much
lower than expected from previous studies. Blue galaxies within the
virial cones of clusters are more prone to projection effects than red
galaxies: \cite{MMR11} analyzed clusters and their member galaxies in
the SDSS, using los velocities and cosmological simulations to
quantify the projection effects. They conclude that $44\pm2\%$ of
galaxies with recent (or ongoing) starbursts that are within the
virial cone are outside the virial sphere. Since galaxies with recent
star formation have blue colours, our BLUE sample includes this
recent-starburst subsample, plus perhaps some more galaxies with more
moderate recent star formation. Moreover, an analysis of cosmological
simulations by \cite{mamon2010} indicates that there is a high cosmic
variance in the fraction of interlopers within the DM particles inside
the virial cone. This suggests that the unusually low concentration of
the blue galaxy sample could be a sign of an unusually high level of
velocity interlopers with low rest--frame velocities in front of and
behind Abell 2142.

\cite{wojtak2010} find a virial radius that corresponds to $\rtwo =
2.15^{+0.10}_{-0.12}\, \Mpc$, in excellent agreement with our
different estimates of the virial radius (Table \ref{Tab: all}). On
the other hand, they find a mass scale radius $r_s=1.0^{+0.3}_{-0.2}\,
\Mpc$ not compatible with our value of the \concordance model,
although in agreement with the results of the DK, MAMPOSSt, and X-ray
analyses. \citeauthor{wojtak2010} assumed that the DM and galaxy scale
radii were equal. Such an unverified assumption may have biased high
their scale radius for the mass distribution.  On the other hand, the
values of the mass scale radii that we found from DK and MAMPOSSt
(0.93 and 0.83 Mpc, respectively, see Table~\ref{Tab: all}) are
consistent with that of the RED galaxy population used as the tracer
(0.95 Mpc), as is in \citeauthor{wojtak2010}, both within the
uncertainties.

\subsection{Velocity anisotropy profile}
The velocity anisotropy profile for the ALL sample in the centre is
compatible with the one found by \cite{wojtak2010}. In the outer part,
at $\simeq 3\, \Mpc$, the value of $\sigma_r/\sigma_\theta$ found by
\cite{wojtak2010} is higher and offset from ours by $1.4\,\sigma$. An
analysis of a stacked sample of 107 nearby ENACS clusters
\citep{biviano2004,katgert2004} shows that the orbits of ellipticals
and S0s (hence red) galaxies are compatible with isotropy, while those
of early and late-type spirals have radial anisotropy. At slightly
higher redshifts, \cite{vandermarel2000} also find red galaxies close
to isotropy. The velocity anisotropy profile for our BLUE sample
presents behaviour that lies in between the profiles found in
\citeauthor{biviano2004} for the early spirals and the late spirals
together with emission line galaxies, suggesting agreement between
their findings and ours. The anisotropy profile we found for the ALL
sample appears to be consistent with those measured by
\cite{lemze2012} and \cite{mamon2013} in simulated $\Lambda$CDM
haloes. In simulations, data are usually stacked or averaged, and the
scatter in the anisotropy profiles is considerable \citep[see
  e.g.][]{lemze2012,mamon2013} and this reflects the variety of
configurations of galaxy clusters. A2142 does not present strong
deviations from the general trend, because its anisotropy profile is
compatible with this scatter.

\subsection{PPSD profile and $\beta-\gamma$ relation}
\cite{bivianoclash} analyzed the pseudo phase--space density on
MACS1206, a cluster at $z = 0.44$. They find a $Q(r)$ profile with a
slope for the blue galaxies in agreement with the predictions of
\cite{dehnen2005}, at odds with our findings. We speculate that this
different behaviour might provide a hint of the dynamical history of
clusters. In fact, a cluster that has only recently undergone the
phase of violent relaxation might present a population of blue
galaxies in rough dynamical equilibrium. On the other hand, a cluster
that has undergone the violent relaxation phase a long time ago,
should have had time to transform its blue galaxies into red
ones. Therefore the blue galaxy population would be mainly composed of
only recently accreted galaxies, hence not in dynamical equilibrium.

While galaxies are biased tracers of the DM velocity dispersion
\citep{munari2013}, if the velocity dispersion profile of the galaxy
component is proportional to that of the DM component at all radii
(i.e. no velocity bias relative to the DM), then the PPSDs built from
the galaxies should have the same slope as the one built from the DM. On
the other hand, if the velocity bias of the galaxy component is a
function of radius, as found by \cite{wu2013} in cosmological
hydrodynamical simulations of clusters, then the PPSD built from the
galaxies will be different from the one built with the DM component
(after proper normalization). Since the DM component dominates the
gravitational potential of clusters, we infer that the consistency of
the PPSD, built with the total density and the velocity dispersion of
the RED galaxy component, suggests that the velocity bias of the
component of red galaxies outside of substructures is roughly
independent of radius.

At all radii, the RED galaxy sample shows somewhat lower $\beta$ for
given $\gamma$ (measured with total mass density) than found in
simulated haloes. However, the $\beta-\gamma$ relations have been
derived using DM-only simulations, which do not take the effects of
the presence of baryons into account. Now, if the tangential and
radial components of the velocity dispersion of the galaxy population
are proportional to those of the DM, then the velocity anisotropy of
the galaxy population, written as ${\cal A}=\sigma_r/\sigma_\theta$
should be proportional to that of the DM, but the non-linear function
of ${\cal A}$, $\beta=1-1/{\cal A}^2$, measured for the galaxies, will
not necessarily be proportional to the analogous $\beta$ for the
DM. Therefore, any radial variation of ${\cal A}$, hence $\beta$, will
lead to a bias in the $\beta-\gamma$ relation.  Finally, the $\beta -
\gamma$ relation may vary from cluster to cluster \citep{ludlow2011}.

\section{Conclusions}
\label{conclusions}

We have computed the mass and velocity anisotropy profiles of A2142, a
nearby ($z = 0.09$) cluster, using the kinematics of cluster
galaxies. After a membership algorithm was applied, we considered the
sample made of all members (ALL sample), as well as two subsamples,
consisting of blue member galaxies (BLUE sample) and red member
galaxies that do not belong to substructures (RED sample).

We made use of three methods based on the kinematics of galaxies in
spherical clusters: DK, MAMPOSSt and Caustic (see
Sect.~\ref{sec:techniques}). The mass profiles, as well as the virial
values of the mass and the radius, are consistent among the different
methods, and they agree with the results coming from the X-ray
\citep{akamatsu2011} and the weak lensing \citep{umetsu2009}
analyses. \cite{serra2011} find that the caustic technique tends to
overestimate the value of mass in the central region of a cluster. Our
results appear consistent with this finding, because the caustic mass
profile increases more rapidly with radius in the inner part with
respect to the profiles coming from DK and MAMPOSSt.

The parameters describing the mass profile are then used to invert the
Jeans equation and compute the velocity anisotropy for the three
different samples considered. Despite large uncertainties, the
$\beta(r)$ profile for the full set of cluster members is compatible
with isotropy, becoming weakly radially anisotropic in the outer
regions. The behaviour of the RED sample is different. Although
compatible within $1\,\sigma$ with isotropy at all radii within
$\rtwo$, it is suggestive of a decreasing slope, starting slightly
radially anisotropic in the centre and becoming slightly tangentially
anisotropic at large radii. The difference between the $\beta(r)$
profiles for the ALL sample and the RED sample is mainly due to the
behaviour of the BLUE sample, which shows radial anisotropy at all
radii except in the centre where it is isotropic.

With the information obtained on A2142, we were able to test some
theoretical relations regarding the interplay between the mass
distribution and the internal kinematics of a cluster. We investigated
the radial profile of the pseudo phase--space density (PPSD) $Q(r)$,
as well as its radial counterpart $Q_r(r)$. When we considered the
total density profile to compute $Q$ and $Q_r$, we found that the
profiles for A2142 are weakly consistent with the theoretical
expectations \citep{dehnen2005,ludlow2010} when considering the ALL
sample, but a good agreement is observed in the RED sample. This
strengthens the scenario of blue galaxies being a population of
galaxies recently fallen into clusters, which have had no time to
reach an equilibrium configuration yet, or are heavily contaminated by
interlopers.

We estimated the PPSD profile of the total matter, making the
assumption that the galaxy velocity dispersion is a good proxy for the
total matter dynamics. \erratum{The PPSDs computed replacing the total
  mass density by the number density of the tracer for which we
  compute the velocity dispersion are consistent with those computed
  with the mass density profile.}

The velocity anisotropy configuration of the internal kinematics
reflects the formation history of the cluster. Therefore we expect a
relation between the velocity anisotropy and the potential of the
cluster. A relation linking the $\beta(r)$ profile and $\gamma(r)$,
the logarithmic slope of the potential, has been analysed and compared
to the theoretical results provided by \cite{hansen2006}, resulting in
weak agreement. A correlation between the $\beta$ and $\gamma$ appears
to hold out to $\gamma \simeq -2.3$ in the RED sample, corresponding
to a radial distance $\simeq 0.5\, r_{200} \simeq 1 \,
\Mpc$. Interestingly, cluster-mass simulated $\Lambda$CDM haloes also
follow the \citeauthor{hansen2006} relation out to slopes of $\gamma
\approx -2.3$ but not beyond (see Fig.~17 of \citealp{lemze2012}).
Our considerations do not change when we compute the $\beta - \gamma$
relation using the logarithmic slope of the number density profile of
galaxies instead of the total matter density profile.

Before reaching any conclusion, we must keep in mind that the present
theoretical studies of the $\beta-\gamma$ and PPSD relations lack the
influence of baryonic physics, as well as the dynamical processes
acting on galaxies but not on DM particles. This might induce the
differences when comparing the theoretical predictions with the
observational results.

When we have better control of these properties, the PPSD might
provide a powerful tool for the study of structure formation. As an
example, the PPSD of the blue galaxies in A2142 appears very different
from what has been found for the blue galaxies in another cluster,
MACS~J1206.2--0847 at $z=0.44$ \citep{bivianoclash}. This discrepancy
suggests interesting perspectives for understanding the formation of
galaxy clusters.

\begin{acknowledgements}
We thank Luca Ciotti, Colin Norman, Barbara Sartoris, and Radek Wojtak
for useful discussions, an anonymous referee for helpful comments, and
Anthony~Lewis for building the public CosmoMC Markov~Chain Monte~Carlo
code. We acknowledge partial support from ``Consorzio per la Fisica -
Trieste'' and from MIUR PRIN2010--2011 (J91J12000450001). AB and EM
acknowledge the hospitality of the Institut d'Astrophysique de Paris.
\end{acknowledgements}

\bibliographystyle{aa} 
\bibliography{bibliography} 
\end{document}